\documentclass[num-refs]{wiley-article}
\usepackage{siunitx}

\papertype{Original Article}
\paperfield{Journal Section}

\usepackage{amsmath}
\usepackage{enumitem}
\usepackage{float}
\usepackage{graphicx}
\usepackage[table]{xcolor}
\usepackage{ragged2e}
\usepackage{booktabs}

\usepackage{pifont}% http://ctan.org/pkg/pifont
\newcommand{\cmark}{\ding{51}}%
\newcommand{\xmark}{\ding{55}}%

\definecolor{darkred}{rgb}{0.5,0,0}     
\definecolor{darkgreen}{rgb}{0,0.5,0}   
\definecolor{darkblue}{rgb}{0,0,0.5} 

\usepackage[colorlinks,bookmarksnumbered,linkcolor=darkblue,citecolor=darkred,urlcolor=darkgreen]{hyperref}

\newcommand{\secref}[1]{Section~\ref{sec:#1}}
\newcommand{\tableref}[1]{Table~\ref{tab:#1}}
\newcommand{\figref}[1]{Figure~\ref{fig:#1}}
\newcommand{\eqaref}[1]{Equation~\eqref{eq:#1}}

\title{The Threat to the Validity of Predictive Mutation Testing: The Impact of Uncovered Mutants}

\author[1]{Alireza~Aghamohammadi}
\author[1]{Seyed-Hassan Mirian-Hosseinabadi}

\affil[1]{A. Aghamohammadi and S. Mirian-Hosseinabadi are with the Department of Computer Engineering, Sharif University of Technology, Iran.}

\corraddress{S. Mirian-Hosseinabadi, Department of Computer Engineering, Sharif University of Technology, Iran.}
\corremail{hmirian@sharif.edu}

\begin{document}

\maketitle

\begin{abstract}

Predictive Mutation Testing (PMT) is a technique to predict whether a mutant will be killed by using machine learning approaches.
Researchers have proposed various machine learning methods for PMT under the cross-project setting.
However, they did not consider the impact of uncovered mutants. A mutant is uncovered if the statement on which the mutant is generated is not executed by any test cases.
 We show that uncovered mutants inflate previous PMT results. Moreover, we aim at proposing an alternative approach to improve PMT and suggesting a different interpretation for cross-project PMT.
 We replicated the previous PMT research. We also proposed an approach based on the combination of Random Forest and Gradient Boosting to improve the PMT results. We empirically evaluated our approach on the same 654 Java projects provided by the previous PMT literature. 
Our results indicate that the performance of PMT drastically decreases in terms of AUC from  0.83 to 0.51. Furthermore, PMT performs worse than random guesses on 27\% of the test projects. The proposed approach improves the PMT results by achieving the average AUC value of 0.61.
\end{abstract}

\section{Introduction}\label{sec:introduction}
Mutation
testing is an approach, which generates several faulty versions of a program by means of applying syntactic changes to source code~\cite{Cambridge-2016-Introduction-Software-Testing,TSE_2011_Jia_An-analysis-and-survey-of-the-development-of-mutation-testing,IST_2011_Offutt_A-mutation-carol}. Each faulty version of the program is called a \textit{mutant}. Mutation testing is generally used to assess the effectiveness of the test suite~\cite{2015-QRS-How-Effective-Code-Coverage,ICSE-2014-Code-Coverage-Suite-Evaluation,TOSEM-2015-Guidelines-for-Comparisions-Tests}. Every test case is executed against each mutant. A mutant is said to be \textit{killed} if there is at least one test case that fails. A mutant is considered as the \textit{survived} mutant if all the test cases pass. The ratio of the number of killed mutants to the total number of non-equivalent mutants is the mutation score of the test suite~\cite{STVR_2018_Zhu_A-systematic-literature-review-of-how-mutation-testing-supports-quality}.

Despite the benefits of mutation testing, it has one major problem~\cite{JSS_2019_Pizzoleto_A-systematic-literature-review-of-techniques-and-metrics-to-reduce-mutation,ICSE_2018_Petrovic_state-of-mutation-testing-at-google} --- a plethora of generated mutants. Therefore, mutation testing is computationally expensive. Researchers have proposed various techniques to reduce the cost of mutation testing such as decreasing the number of executed test cases~\cite{ICST_2018_Gopinath_If-you-cant-kill-a-supermutant-you-have-a-problem,ICSE_2016_Devroey_Featured-Mode-based-mutation-analysis,JSS_2016_Ma_Mutation-testing-cost-reduction-by-clustering-overlapped-mutants}, limiting the creation of specific mutants~\cite{ICST_2017_Iida_Reducing-mutants-with-mutant-killable-precondition,GPCE_2017_Fernandes_Avoiding-useless-mutants,STVR_2015_Just-Higher-accuracy-and-lower-run-time-efficient-mutation-analysis-using-non-redundant-mutation-operators,ICSE_2010_Steimann_From-behaviour-preservation-to-behaviour-modification}, selecting certain types of mutants~\cite{ICSE_1993_Offutt_An_experimental-evaluation-of-selective-mutation,ESE_2020_Chekam_Selecting-fault-revealing-mutants}, and predicting mutant execution results~\cite{TSE_2019_Zhang_Predictive-Mutation-Testing,ICST_2019_Mao_An-Extensive-Study-on-Cross-Project-Predictive-Mutation-Testing}.

Predicting mutant execution results is a method to predict whether a mutant will be killed by means of using machine learning techniques. Zhang et al.~\cite{TSE_2019_Zhang_Predictive-Mutation-Testing} introduced a new dimension of mutation testing. They proposed an approach named Predictive Mutation Testing (PMT), which predicts mutant execution results by using software metrics as input to the machine learning techniques. Mao et al.~\cite{ICST_2019_Mao_An-Extensive-Study-on-Cross-Project-Predictive-Mutation-Testing} extended Zhang et al.'s work by adding a number of projects and performing PMT under the \textit{cross-project} setting. The cross-project PMT exploits a machine learning model to predict mutant execution results for totally unseen new projects. 

However, Zhang et al.'s work and Mao et al.'s work ~\cite{TSE_2019_Zhang_Predictive-Mutation-Testing,ICST_2019_Mao_An-Extensive-Study-on-Cross-Project-Predictive-Mutation-Testing} both have an assumption that may inflate their results. We call this assumption \textit{uncovered mutants}. A mutant is uncovered if the statement on which the mutant is generated is not executed by any test cases~\cite{Just2011,Mateo2014}. As uncovered mutants certainly will be survived, there is no point in learning such mutants. Uncovered mutants could bear on the results of PMT. We discuss the details of PMT in \secref{Related-work}.

The main purpose of our research is to show that uncovered mutants can inflate the results of PMT. We also propose an approach to improve the results of PMT. To this end, we pose three research questions:

\begin{description}[style=unboxed,leftmargin=0cm]
\item[RQ1:] To what extent the uncovered mutants affect the results of Predictive Mutation Testing?
\\
RQ1 tries to investigate the effects of uncovered mutants on the previous research results. We hypothesize that this assumption could drop the performance of the machine learning models significantly. Indeed, we show that in the $27\%$ of the test projects, the previous models perform worse than random guesses. Then, we propose our approach to improve PMT. The next research question assesses the quality of our proposed approach.
\item[RQ2:] How well the proposed approach does perform compared to state-of-the-art approaches?
\\
RQ2 assesses our proposed approach in comparison to other PMT methods. In this research question, we empirically show that the proposed approach outperforms the previous approaches in terms of unbiased metrics for imbalanced data such as Area Under the ROC Curve (AUC), Balanced Accuracy, and Matthews Correlation Coefficient (MCC).
\item[RQ3:] Which features are important in predicting the execution results of mutants?
\\
In RQ3, we would like to capture the most important features  in predicting mutant execution results. In particular, we would seek any difference in the interpretation of PMT before and after considering uncovered mutants. If we could show that the ranks of features in predicting such mutants significantly change, it signifies that the prior research interpretations are misleading.
\end{description}

In summary, our contributions are as follows:
\begin{itemize}
\item We investigate the impact of uncovered mutants on the results of Predictive Mutation Testing. In particular, we show that the performance of previous research decreases significantly from $0.83$ to $0.51$ in terms of AUC.

\item We propose an approach based on Random Forest and Gradient Boosting by considering the effect of uncovered mutants and improve the results of Predictive Mutation Testing in terms of various unbiased metrics (towards imbalanced data) such as AUC, MCC, and Balanced Accuracy.

\item We examine the most important features in predicting the execution results of mutants and find some contradiction when we consider the impact of uncovered mutants.

\item Similar to Mao et al.'s work and Zhang et al.'s work~\cite{TSE_2019_Zhang_Predictive-Mutation-Testing, ICST_2019_Mao_An-Extensive-Study-on-Cross-Project-Predictive-Mutation-Testing}, we make all our source code publicly available for other researchers to further investigate and improve the Predictive Mutation Testing.
\end{itemize}

The rest of the paper is organized as follows. In \secref{Related-work}, we elaborate upon the related work as well as provide background information to understand PMT. \secref{RQ1} investigates the impact of uncovered mutants on the results of PMT. In \secref{Proposed-Approach}, we detail our proposed approach to improve PMT by considering the effects of uncovered mutants. \secref{RQ3} focuses on the interpretation of the proposed approach and the difference between our interpretation and previous research. In \secref{Results}, we express the results of our research questions. 
\secref{Threats-to-Validity} reviews threats to the validity of this paper. We draw conclusions and present our future plans in \secref{Conclusions}.

\section{Related Work and Background}\label{sec:Related-work}
In this section, we provide the necessary information to understand our paper and also review the related work.

Zhang et al.~\cite{TSE_2019_Zhang_Predictive-Mutation-Testing} introduced the notion of Predictive Mutation Testing (PMT). PMT is an approach based on machine learning techniques that predicts the execution results of mutants, namely killed or survived, without conducting mutation testing. The idea is anticipating the mutant execution results beforehand to reduce the overhead of mutation testing. Zhang et al.~\cite{TSE_2019_Zhang_Predictive-Mutation-Testing} collected 163 real-world Java projects and demonstrated that the PMT obtained the Area Under the ROC Curve (AUC) value of 0.80 on average. Based on their results,
Random Forest outperformed other machine learning models in predicting mutant execution results.
They defined 16 different metrics for each project. These metrics can be categorized as either static or dynamic features. The dynamic features such as \texttt{numExecuted} and \texttt{numTestCover} are related to the run-time behavior of the program. The first one determines the number of times each mutant is executed. The latter indicates how many test cases execute the mutant. The static features such as \texttt{infoComplexity} and \texttt{LOC} are regarding the compile-time information of source code. Interested readers can refer to the original paper to see the full list of feature names accompanied by definition~\cite{TSE_2019_Zhang_Predictive-Mutation-Testing}.

Mao et al.~\cite{ICST_2019_Mao_An-Extensive-Study-on-Cross-Project-Predictive-Mutation-Testing}
extended PMT by introducing 654 real-word Java projects and 95 features. Their data set and features cover the Zhang et al.'s work~\cite{TSE_2019_Zhang_Predictive-Mutation-Testing}.
As well as two preivious dynamic features, they presented two other dynamic ones, namely \texttt{numAssertInTM} and \texttt{numAssertInTC}. The first one refers to the number of assertions in test methods that exercise the mutant. The latter indicates the number of assertions in test classes that exercise the mutant. They also gathered two categorical features, namely \texttt{MutatorClass} and \texttt{returnType}. \texttt{MutatorClass} shows the type of mutation operator. \texttt{returnType} refers to the return type of the method in which the mutant is located. Interested readers could study the original paper to see the full name of features as well as the definition of each one~\cite{ICST_2019_Mao_An-Extensive-Study-on-Cross-Project-Predictive-Mutation-Testing}. Mao et al. trained various machine learning models to assess the PMT under the cross-project settings.
They also considered other models such as deep learning approaches. Similar to Zhang et al.'s work~\cite{TSE_2019_Zhang_Predictive-Mutation-Testing}, Random Forest also outperformed the state-of-the-art techniques in extended version of PMT. They achieved the AUC value of 0.89.

However, there is an assumption in both of the Zhang et al.'s and Mao et al.'s work~\cite{TSE_2019_Zhang_Predictive-Mutation-Testing,ICST_2019_Mao_An-Extensive-Study-on-Cross-Project-Predictive-Mutation-Testing}, which may affect their results. Every mutant that is not executed will be certainly survived; thus,  there is no point in learning such mutants. Previous research did not consider the effect of those mutants, nonetheless, it might reduce the effectiveness of PMT and inflate their results. Our motivation in this paper is to explored the influence of uncovered mutants and further improve the results of PMT by considering non-executed mutants.

Chekam et al.~\cite{ESE_2020_Chekam_Selecting-fault-revealing-mutants} proposed an approach to select mutants that are fault revealing. A mutant is said to be fault revealing if the mutant is not equivalent to the original program and unveils a program fault. They defined 28 static features of the program and introduced a machine learning approach (i.e., Random Forest) to predict such mutants.
They evaluated their approach on Codeflaws and CoREBench \cite{ICSE_2017_Tan_CodeFlaws, ISSTA_2014_CoREBench}, which are the programs written in the C programming language.
They also, for the first time, proposed a concept of mutant prioritization. A mutant with a higher probability of revealing faults has a higher priority for being executed in the mutation prioritization.

\section{RQ1: The Impact of Uncovered Mutants}\label{sec:RQ1}
In this section, we examined the effectiveness of previous research with respect to PMT by considering the impact of uncovered mutants. Every non-executed mutant is certainly survived. Consequently, there is no point in learning such mutants. \tableref{Covered-Uncovered} shows the number of covered and uncovered mutants. On the basis of this table, $62\%$ of the total mutants are not executed, which is significantly more than half of all the mutants available in the data set.  The quantity of non-executed mutants may inflate the results of previous research.

\begin{table}[htpb]
\renewcommand{\arraystretch}{1.3}
  \caption{The number of covered and uncovered mutants in the data set}
  \centering
  \label{tab:Covered-Uncovered}
  \begin{tabular}{@{} cc@{} }
   \toprule
    \bfseries Is Executed? & \bfseries The Number of Mutants\\
   \midrule
    \cmark & 1137336\\
    \xmark& 1894940\\
   \bottomrule
  \end{tabular}
\end{table}

Zhang et al. are the first ones who introduce the concept of PMT~\cite{TSE_2019_Zhang_Predictive-Mutation-Testing}. They showed that Random Forest gained the best results in predicting the execution results of mutants. Mao et al. extended Zhan et al.'s work by increasing the number of projects and features~\cite{ICST_2019_Mao_An-Extensive-Study-on-Cross-Project-Predictive-Mutation-Testing}.
Similarly, Random Forest outperformed other machine learning models in the extended data set.
Mao et al.'s work~\cite{ICST_2019_Mao_An-Extensive-Study-on-Cross-Project-Predictive-Mutation-Testing} improved the previous research and subsumed the data set and independent features. 
To investigate the impact of uncovered mutants, we replicated the same procedure described by Mao's work~\cite{ICST_2019_Mao_An-Extensive-Study-on-Cross-Project-Predictive-Mutation-Testing}.
We used the same data set and the same machine learning model (i.e., Random Forest) to explore the effectiveness of PMT.
Specifically, we randomly chose 66 projects out of 654 projects as the test data and investigated the performance of PMT in predicting covered mutants to unveil any inflation that uncovered mutants might produce.

\section{RQ2: Proposed Approach}\label{sec:Proposed-Approach}
\begin{figure*}[htpb]
\centering
\includegraphics[width=0.95\textwidth]{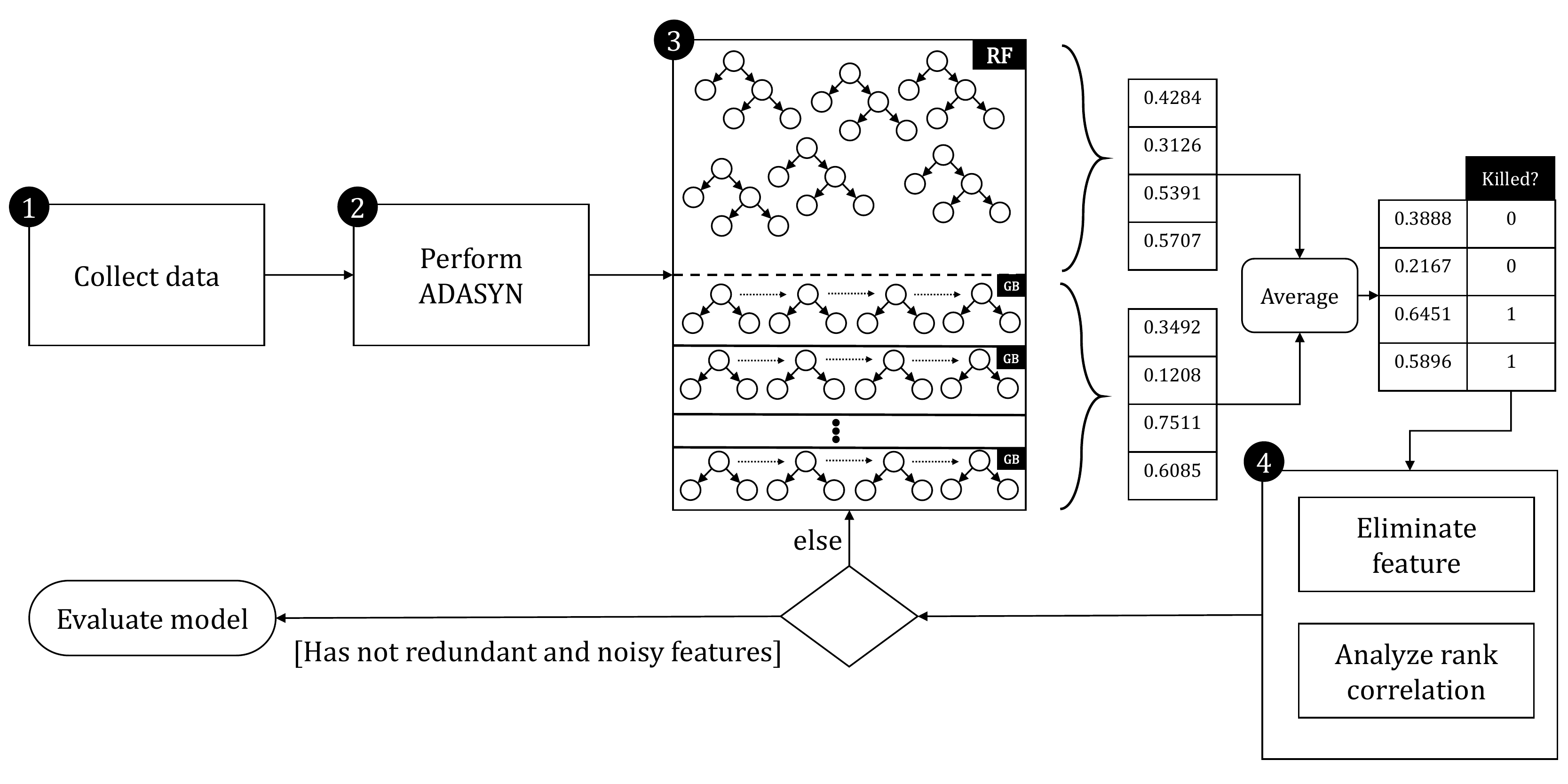}
\caption{Our high-level proposed approach}
\label{fig:Propsed-Approach}
\end{figure*}

Here, we elaborate upon the proposed approach.
\figref{Propsed-Approach} illustrates our high-level proposed approach.

First, we gathered data (Step 1). We used the same data introduced by Mao et al.~\cite{ICST_2019_Mao_An-Extensive-Study-on-Cross-Project-Predictive-Mutation-Testing}. As suggested by other researchers and practitioners~\cite{Ripley1996,James2013,Russell2010}, we split our data into three independent sets, namely train, validation, and test. 
Specifically, 522 projects were used in the train part, 66 projects were located in the validation set, and 66 projects were utilized for the final evaluation process. It is worth mentioning that any hyperparameter tuning was performed based on the result of the machine learning model on the validation set. Indeed, the test set did not participate in the hyperparameter tuning process due to avoid bias in the evaluation part.
 Then, because of the imbalanced nature of the data set, we used Adaptive Synthetic Sampling Approach (ADASYN) to rebalance the data set (Step 2)~\cite{IJCNN_2008_He_ADASYN}. By doing so, we ensured that the number of killed and survived mutants are equal in the train set.
 
 Afterward, we devised a predictive model based on the combination of Random Forest and Gradient Boosting (Step 3). In particular, our Random Forest consisted of 100 Decision Trees. Also, we used the combination of 50 Gradient Boosting to reduce variance. We averaged the prediction of Random Forest and Gradient Boosting as a final prediction of the model.
Because the data may have some noisy and redundant features, we preprocessed data to remove them recursively (Step 4). Previous research shows that eliminating noisy and highly correlated features would improve the results of the machine learning model. To this end, we employed the recursive feature elimination technique along with rank correlation analysis to acquire the set of effective features in predicting the execution results of mutants. In particular, we removed the least important feature that has more than 0.9 Spearman's rank correlation with another feature. Then, we tuned the hyperparameters of the model on the validation set. We had carried out Step 3 and Step 4 recursively until all of the highly correlated and noisy features were eliminated.
 Ultimately, the final trained model was evaluated on the test set. In the following, we detail each step of the proposed approach.
 
\subsection{Step 1: Collect data}
Zhang et al.~\cite{TSE_2019_Zhang_Predictive-Mutation-Testing} introduced 163 real-world Java projects in order to predict the execution results of mutants. Afterward, Mao et al.~\cite{ICST_2019_Mao_An-Extensive-Study-on-Cross-Project-Predictive-Mutation-Testing} extended Zhang et al.'s work and provided 654 real-world Java projects. In this paper, we used the same 654 Java projects presented by Mao et al.~\cite{ICST_2019_Mao_An-Extensive-Study-on-Cross-Project-Predictive-Mutation-Testing}. We split the data into three sets, namely the train ($80\%$), validation ($10\%$), and test ($10\%$) set.
In particular, we randomly chose 522 Java projects for the train set, 66 Java projects for the validation set, and 66 Java projects for the test set. 
To avoid inflating results, we removed uncovered mutants and considered only those mutants to be executed at least once. 
The validation set would be used to tune model hyperparameters. Note that the test set was not used for tuning hyperparameters. After finding all the hyperparameters, we utilized the test set to evaluate the machine learning model. \tableref{Train-Valid-Test} presents some statistical information with respect to the train, validation, and test set.

\begin{table}[htpb]
\renewcommand{\arraystretch}{1.3}
\centering
  \caption{Statistical information about the train, validation, and test set}
  \label{tab:Train-Valid-Test}
  \begin{tabular}{@{} lccc@{} }
   \toprule
    &\bfseries Train&\bfseries Validation&\bfseries Test\\
   \midrule
    \# Mutants& 1011364 &67496 &58476\\
    \# Killed& 680474 & 50839 & 37843\\
    \# Survived& 330890 & 16657 & 20633\\
   \bottomrule
  \end{tabular}
\end{table}

\subsection{Step 2: Perform ADASYN}
One of the challenges in today's machine learning problems is to tackle with imbalanced data sets~\cite{TR_2013_Wang-Using-Class-Imbalance-Learning-for-Software-Defect-Prediction, ESWA_2017_Guo_Learning-from-class-imbalanced-data,TSE_2018_Tantithamthavorn_The-Impace-of-class-rebalancing-techniques-on-the-preformance, ESE_2017_Malhotra_An-empricial-study-for-software-change-prediction-using-imbalanced-data, ICTAI_2010_Khoshgoftaar_Attribute-selection-and-imbalanced-data}. A data set is said to be imbalanced if one of the dependent variables dominates the sample space. Over the course of years, researchers have proposed a plethora of approaches to solve the issue of imbalanced data such as randomly oversampling with replacement, randomly under-sampling with replacement, and synthetic minority oversampling technique~\cite{JAIR_2002_SMOTE}. Besides, rebalancing techniques have been shown that improve the performance of prediction models in terms of AUC and recall~\cite{TSE_2018_Tantithamthavorn_The-Impace-of-class-rebalancing-techniques-on-the-preformance}.

In this paper, we used a famous approach named Adaptive Synthetic Sampling Approach (ADASYN) to handle our imbalanced data~
\cite{IJCNN_2008_He_ADASYN}. ADASYN lies on the oversampling techniques. It oversamples minority class synthetically while considering the difficulty of those minor instances. The instances that belong to the minority class will be more generated if they are harder to learn. ADASYN has the advantage of changing the decision boundary between difficult instances and easy ones. It is worth to note that we just rebalanced the train data and did not touch the validation and test set. The validation and test set are aimed at representing unseen data.

\subsection{Set 3: Train the Model}
Our machine learning model has two folds. The first fold is Random Forest.
Random Forest is a \textit{bagging} approach, which tries to train different decision trees on data~\cite{RF_2001}. Each tree is trained on a different portion of the data. Random Forest aggregates the results:

\begin{equation}\label{eq:RF}
    \mathrm{RF}\left (X\right ) =  \sum_{i=1}^{N}\frac{1}{N}\mathrm{DT_i}\left (X\right )
\end{equation}

In \eqaref{RF},  $\mathrm{RF}$ is determined based on the average of $N$ different $\mathrm{DT_i}$ scores --- $N$ is the number of Decision Trees. One advantage of Random Forest is reducing the variance of a predictor (i.e., Decision Tree)~\cite{Book_20120_Murphy_Machine-Learning}.

The second fold is Gradient Boosting. Boosting is an approach based on the combination of \textit{weak learners}~\cite{Book_2014_BenDavid}. A weak learner is an estimator that is slightly better than a random guess. For example, Decision Tree with the depth of one, Decision Stump, is a weak learner. Boosting makes a decision based on aggregating weak learners:

\begin{equation}\label{eq:GB}
        \mathrm{GB}\left (X\right ) =  \sum_{i=1}^{T}\alpha_i\mathrm{WL_i}\left (X\right )
\end{equation}

In \eqaref{GB}, $\mathrm{GB}$ is calculated based on the decision of $T$ weak learners $\mathrm{WL}$ --- $\alpha_i$ is the weight attributed to $\mathrm{WL_i}$. One limitation of Gradient Boosting is overfitting~\cite{Book_2018_Mohri}. To avoid overfitting, we used bagging. In other words, we created multiple Gradient Boostings and trained them on a different subset of the data. We averaged the results to reduce the variance.

Finally, for each data point (each mutant), we obtained two predictions. The first one is the estimation of Random Forest. The second one is the prediction of bagging of different Gradient Boostings. Our final prediction for each mutant is the mean of those two values. If the prediction is greater than $0.5$, we predict the mutant would be killed. If the prediction is less than $0.5$, we predict the mutant would be survived.
\subsection{Step 4: Eliminate Features and Analyze Rank
Correlation}
Noisy features could decrease the performance of the model. We consider a feature as noisy if it has less than $1\%$ \textit{importance} with respect to other features in predicting dependent variables. To obtain the importance of each feature, we used the \textit{permutation importance}~\cite{SC_1017_Gregorutti-Correlation-and-variable-importance-in-random-forests,BI_2010_Altmann_Permutation-importance}. The permutation importance is an iterative technique that in each iteration, one of the independent variable values is shuffled --- with respect to the original data --- decorrelating it with the dependent variable. If the performance of the model is improved, it signifies that the feature is noisy, and we remove it. Otherwise, we record the amount of reduction.
These numbers show the importance of the features.
It is worth mentioning that we assessed the performance of the model based on AUC.
We removed the feature with less than $1\%$ importance with respect to other independent variables. Note that we used the \textit{Recursive Feature Elimination} approach~\cite{ML_2002_Guyon_Gene-Selection-for-Cancer-Classification}. We had removed the noisy feature and carried out the permutation importance process recursively until all of the features had importance more than $1\%$ with respect to other features. 

\begin{figure*}[htpb]
\centering
\includegraphics[width=\textwidth]{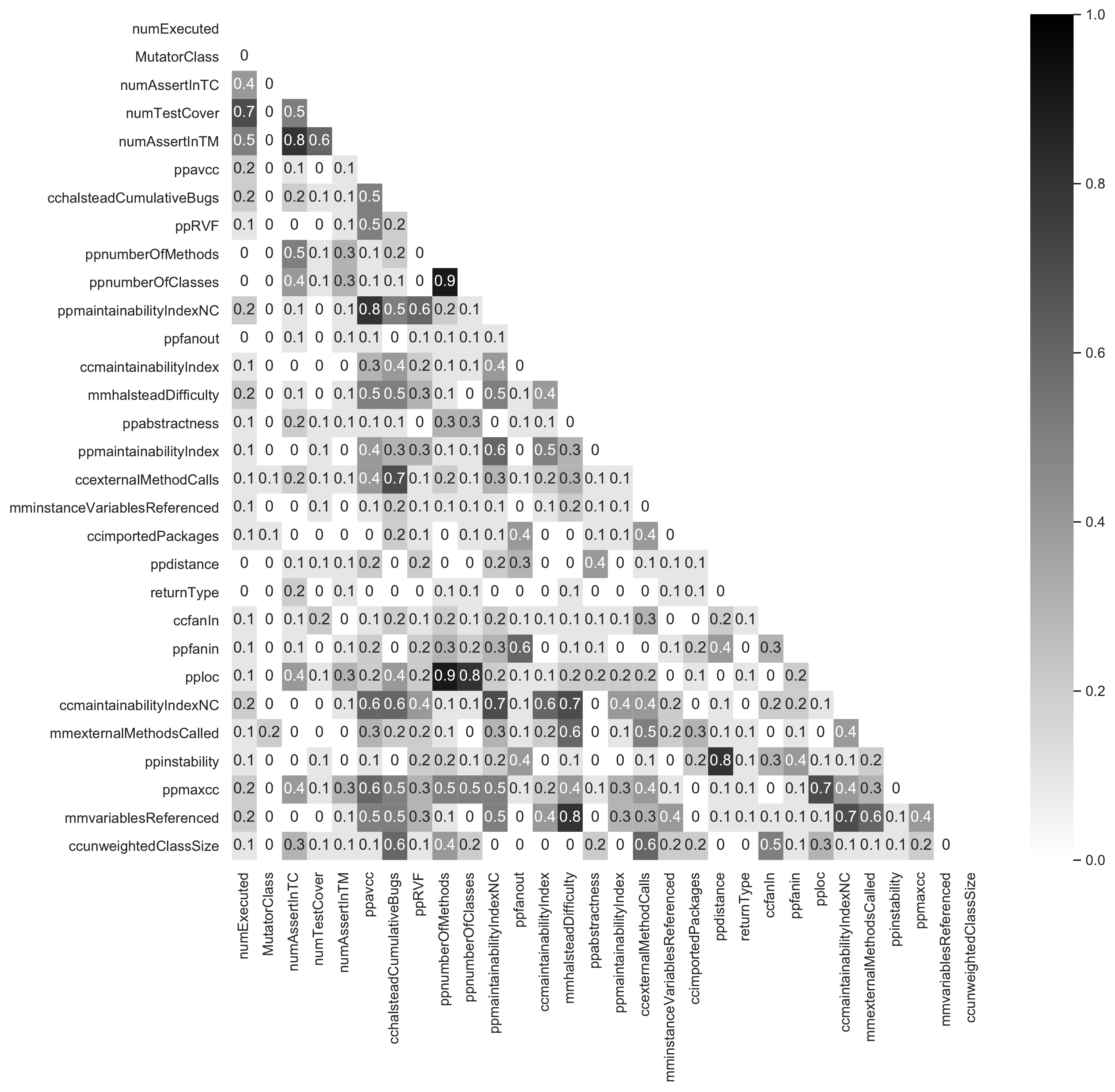}
\caption{Remaining 30 features accompanied by the Spearman's Rho rank correlation}
\label{fig:Rank-Correlation}
\end{figure*}

Redundant features could also reduce the performance of the model. Our machine learning model is a tree-based approach. In the tree-based model, the absolute values of one feature are not important, but the relative values are crucial. We exploited Spearman's Rho rank correlation~\cite{SpearmansRho}. Spearman's Rho is a non-parametric rank correlation metric, which measures the rank correlation between two variables. We consider a feature F\textsubscript{1} as redundant if there is another feature F\textsubscript{2} in the data having the rank correlation of more than $0.9$ with F\textsubscript{1}. In this case, we removed the feature with less permutation importance. After removing noisy and redundant features, we obtained 30 features. \figref{Rank-Correlation} presents the remaining 30 features along with Spearman's Rho rank correlations.  

\subsection*{Implementation Details}

We used the \texttt{Pandas} library for reading and writing data~\cite{Pandas}. Our implementation of the machine learning model is based on \texttt{Scikit-Learn}, a powerful Python machine learning framework~\cite{SKlearn}.
There are two category features in the independent variables, namely \texttt{MutatorClass} and \texttt{returnType}. \texttt{MutatorClass} represents the mutation operator. \texttt{returnType} shows the type of the return value of the method. Since our proposed approach is tree-based, it cannot handle categorical features. We transformed those two categorical features into numerical features by exploiting \textit{Frequency Encoding}. Frequency Encoding is an approach that counts the number of values for each category. Then, it maps each category to its frequency.  
We set the number of estimators of Random Forest to 100 (\texttt{n\_estimators=100}). Each estimator in each split considers $0.7 * 30 = 21$ independent variables  (\texttt{max\_features=0.7}). We employed \texttt{HistGradientBoostingClassifier} to implement the Gradient Boosting part. We fitted 50 different Gradient Boostings and averaged their predictions to reduce the variance. 

\section{RQ3: Interpretation}\label{sec:RQ3}
Here, we interpret our prediction model by means of investigating the most important features. Besides, we compare the most important features in the PMT approach~\cite{ICST_2019_Mao_An-Extensive-Study-on-Cross-Project-Predictive-Mutation-Testing} to our features we obtained in the proposed approach. To this end,  we exploited the method called permutation importance to get the most influential features~\cite{BI_2010_Altmann_Permutation-importance}.

Mao et al.~\cite{ICST_2019_Mao_An-Extensive-Study-on-Cross-Project-Predictive-Mutation-Testing} extracted 12 features out of 95 independent variables that have the most effect on the ability of PMT to predict the mutant execution results. Similarly, we extracted 12 most important features out of 30 independent variables before which were introduced in our proposed approach. Then, we compared our 12 features to those 12 features of PMT. We would seek any difference between these two sets of features that may impact the interpretation of ours towards Predictive Mutation Testing.

\section{Results}\label{sec:Results}
In this section, we present our results as well as our own interpretation regarding the three research questions. 

\subsection{RQ1: The Impact of Uncovered Mutants}
For answering RQ1, we computed the Area Under the ROC Curve (AUC). The ROC curve illustrates the relationship between the true positive rate and the false positive rate. AUC is a measure ranging in $\left[0,1\right]$, which is unbiased towards an imbalanced data set. The AUC value of one signifies the perfect predictive model. The AUC value of 0.5 is similar to random guesses. The AUC value of less than 0.5 is worse than random guesses.

\tableref{RQ1-AUC} shows the performance of Random Forest (the best model in the original paper) in terms of AUC on both the whole data set and only covered mutants. Based on the table, the AUC drops significantly on covered mutants; that is, the median AUC value decreases from $0.833$ to $0.516$, which is slightly better than random guesses. It is worth mentioning that 18 out of 66 (i.e., about $27\%$) projects, the model performs worse than random guesses. On average, the performance of the original PMT drastically decreases from $0.814$ to $0.540$ in terms of AUC. It is worth noting that only five projects have the AUC value of more than $0.6$. We elaborate upon the performance of PMT as well as the distribution of performance in the following research question results.

\begin{table}
\renewcommand{\arraystretch}{1.3}
 \centering
  \caption{The performance of PMT by considering uncovered mutants}
  \label{tab:RQ1-AUC}
  \begin{tabular}{@{} ccc@{} }
   \toprule
    &\bfseries Average&\bfseries Median\\
   \midrule
    All the data& $0.814$&$0.833$\\
    Only covered Mutants&$0.540$&$0.516$\\
   \bottomrule
  \end{tabular}
\end{table}

\subsection{RQ2: Proposed Approach}
\begin{figure}[htpb]
\centering
\includegraphics[scale=0.5]{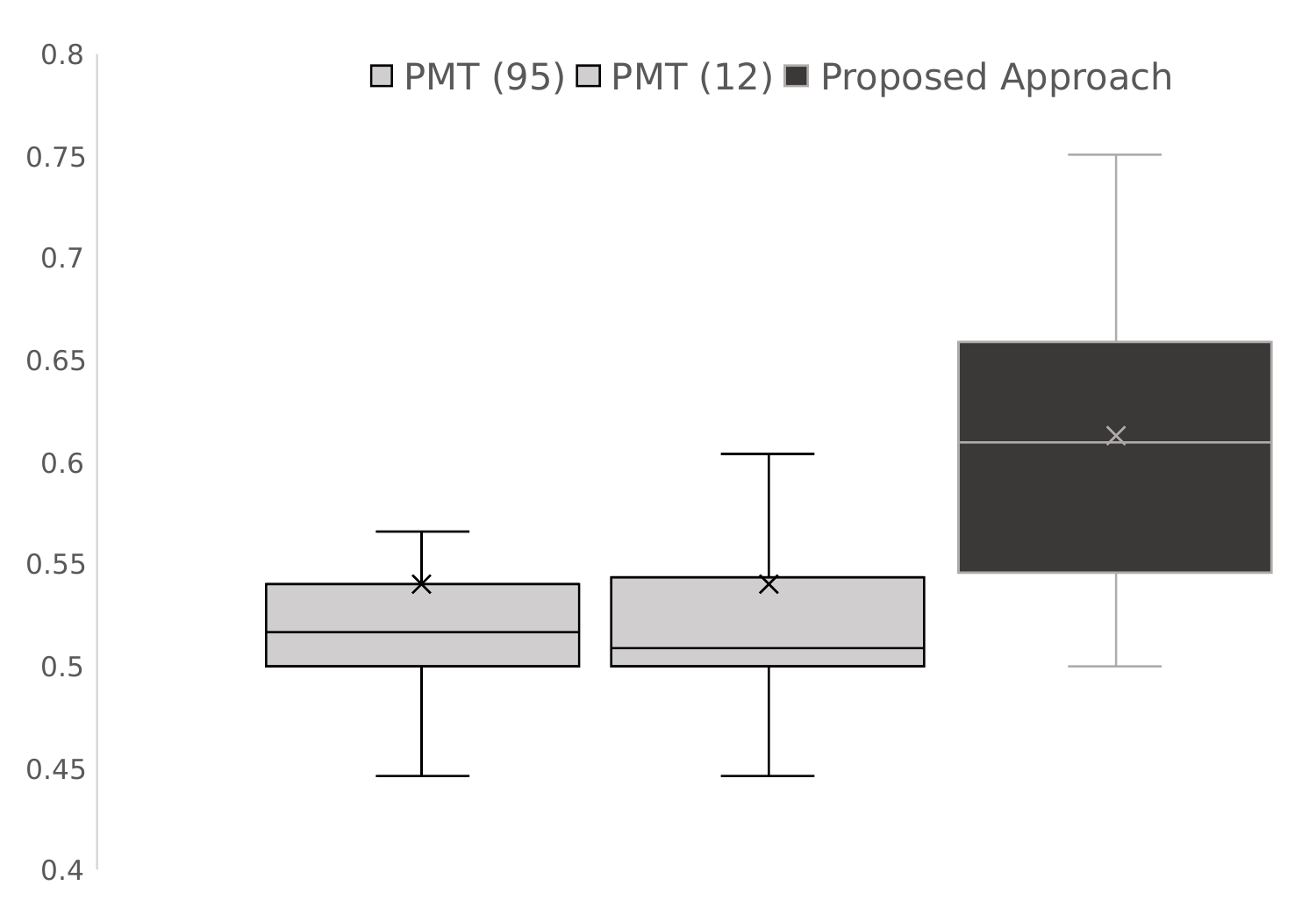}
\caption{The performance distributions in terms of AUC}
\label{fig:BoxPlot_AUC}
\end{figure}

\figref{BoxPlot_AUC} depicts the performance distributions of our proposed approach and original PMT in terms of AUC. The PMT(95) means all the 95 features were used in the training process. The PMT(12) means the top 12 important features were exploited in the training process as suggested by Mao et al.~\cite{ICST_2019_Mao_An-Extensive-Study-on-Cross-Project-Predictive-Mutation-Testing}. Our proposed approach achieved the AUC value of $0.613$ on average. Also, the median AUC value of our model is $0.609$. It indicates about $13\%$ and $20\%$ improvement regarding the average and median of PMT. Unlike PMT(95) and PMT(12), our proposed approach gains the AUC value of more than $0.5$ in all of the test projects meaning no project performs worse than random guesses. It is another improvement compared to PMT.

To better understand the difference between our approach and PMT. We exploited the Scott-Knott Effect Size Difference test~\cite{TSE_2017_Tantithamthavorn_SCOTT,TSE_2019_Tantithamthavorn_SCOTT} with a confidence interval of $0.95$. This test is a variation of the Scott-Knott test that considers Cohen’s delta effect size~\cite{Cohen_delta} and solves the normality assumption of the Scott-Knott test~\cite{SCOTT_KNOTT}. The test ranks the different groups with a statistical significance difference between means.
In \figref{BoxPlot_AUC}, the results of the test are shown in different colors. Our proposed approach achieves the first rank (darker one), yet, PMT(95) and PMT(12) both obtains the second rank (lighter ones). There is no statistical significance difference between PMT(95) and PMT(12).

We also calculated other unbiased metrics that perform well under the imbalanced data set settings to evaluate our proposed approach.
The metrics for imbalanced data are important since other metrics such as precision, recall, and $\mathrm{F}_1$ could be misleading in that condition~\cite{Bio_2011_Powers_Evaluation-from-precition-recall-fmeasure}.

Matthews Correlation Coefficient (MCC) is one of those unbiased metrics~\cite{Bio_2000_Baldi_MCC}. Unlike $\mathrm{F}_1$, MCC takes true negative (TN) into its account~\cite{EASE_2020_Yao_MCC-matters}. The MCC is located in the range of $\left [-1,1\right ]$. The MCC value of one means the perfect predictive model, while the MCC value of zero means the random guesses. Note that the negative MCC value signifies that the model performs worse than random guesses.

Balanced Accuracy is a variation of accuracy which solves the problem of traditional accuracy under the imbalanced  settings~\cite{ICPR_2010_Brodersen_Balanced-Accuracy}. We used the adjusted version of Balanced Accuracy implemented by \texttt{Sklearn}. Similar to MCC, Balanced Accuracy lies in the range of $\left [-1,1\right ]$. The same interpretation holds for this metric.

\tableref{RQ2-Eval} shows the results of the aforementioned metrics. We calculated means of these metric values for both PMT and proposed approach as well as each part of our solution to better understand the participation rate of different ideas. Based on the results, the proposed approach outperforms PMT(95) and PMT(12) in all of the aforementioned  metrics. We improved the Balanced Accuracy value by $88\%$, MCC value by $73\%$ and AUC value by $13\%$. Random Forest with the 30 features, RF(30), obtained in Step 4 of our proposed approach, even outperforms both PMT(95) and PMT(12). When we use ADASYN to balance our data set, the results improve. Based on this table, the combination of ADASYN and 50 Gradient Boostings outperforms the combination of ADASYN and Random Forest. Finally, our proposed approach which exploits ADASYN, Random Forest, and 50 Gradient Boostings, gain the best performance in terms of Balanced Accuracy, MCC, and AUC. These results show that the combination of Bagging and Boosting can improve the results (in the Prediction Mutation Testing settings) significantly.

\begin{table*}[htpb]
\renewcommand{\arraystretch}{1.3}
\centering
  \caption{The performance of proposed approach compared to PMT and each part of our solution}
  \label{tab:RQ2-Eval}
  \begin{tabular}{lccc}
   \toprule
    &\bfseries Balanced Accuracy&\bfseries MCC&\bfseries AUC\\
   \midrule
    PMT(95)&$0.122$&$0.138$&$0.540$\\
    PMT(12)&$0.113$&$0.126$&$0.539$\\
    RF(30)&$0.131$&$0.161$&$0.571$\\
    ADASYN + RF(30)&$0.169$&$0.170$&$0.580$\\
    ADASYN + GBs(30)&$0.205$&$0.209$&$0.605$\\
    Proposed Approach&$\boldsymbol{0.230}$&$\boldsymbol{0.239}$&$\boldsymbol{0.613}$\\
   \bottomrule
  \end{tabular}
\end{table*}

\subsection{RQ3: Interpretation}
\tableref{RQ3-Interpretation} presents the feature importances of PMT(12) and our proposed approach.
The results specify that 7 out of top 12 important features are common to the two approaches, namely \texttt{numAssertInTC} (the number of assertions in the test classes that execute the mutant), \texttt{numExecuted} (the number of execution), \texttt{numTestCover} (the number of tests covering the mutant), \texttt{numAssertInTM} (the number of assertions in the test methods that execute the mutant), \texttt{MutatorClass} (the mutator operator), \texttt{ppavcc} (the package-level average Cyclomatic Complexity~\cite{McCabe1989}), and \texttt{ppmaintainabilityIndexNC} (the package level maintainability index).

However, there are some differences between these two sets.
First, \texttt{numTestCover} drops in ranking from the 3rd to the 12th (the last one in our proposed approach). Second, there are only four features having the feature importance rate of more than $5\%$ in PMT. All of the 12 important features in the proposed approach have the feature importance rate of more than $5\%$, which is statistically non-negligible.  Third, although the dynamic features are the most important ones in PMT (top four), static features in predicting mutant execution results play a crucial role in the proposed approach. For instance, \texttt{mmhalsteadDifficulty}, which is the method-level Halstead difficulty metric~\cite{TSE_1988_Weyuker_Halstead}, is the second most important feature in the proposed approach. Fourth, all the static features apart from \texttt{MutatorClass}, in PMT,  are package-level metrics, however, the static features in the proposed approach comes from a different level of granularity such as method, class, and package level. Finally, some of the static features are quite simple in PMT such as  package lines of code (\texttt{ploc}), the number of package statements (\texttt{ppnumberOfStatements}), and the number of classes in the package (\texttt{ppnumberOfClasses}). However, the static features in the proposed approach are quite complex ones, which measure the relationship between elements of the code, for example, \texttt{mmhalsteadDifficulty} and \texttt{cchalsteadCumulativeBugs}.

\begin{table*}[htpb]
 \renewcommand{\arraystretch}{1.3}
 \centering
  \caption{The top 12 important features in PMT and the proposed approach (PA)}. 
  \label{tab:RQ3-Interpretation}
  \begin{tabular}{cccc}
   \toprule
    \bfseries Feature Names (PMT)&\bfseries Importance (PMT)&\bfseries Feature Names (PA)&\bfseries Importance (PA)\\
   \midrule
    numAssertInTC&$0.233$&numExecuted&$0.165$\\
    numExecuted&$0.222$&mmhalsteadDifficulty&$0.118$\\
    numTestCover&$0.191$&MutatorClass&$0.091$\\
    numAssertInTM&$0.143$&ppavcc&$0.084$\\
    MutatorClass&$0.047$&cchalsteadCumulativeBugs&$0.082$\\
    ppavcc&$0.030$&numAssertInTC&$0.077$\\
    ppmaintainabilityIndexNC&$0.027$&numAssertInTM&$0.076$\\
    ppmaintainabilityIndex&$0.021$&ppnumberOfMethods&$0.065$\\
    ppnumberOfClasses&$0.021$&ppmaintainabilityIndexNC&$0.060$\\
    pptcc&$0.020$&ccmaintainabilityIndex&$0.059$\\
    ppnumberOfStatements&$0.020$&ccmaintainabilityIndexNC&$0.058$\\
    pploc&$0.019$&numTestCover&$0.055$\\
   \bottomrule
  \end{tabular}
\end{table*}

\section{Threats to Validity}\label{sec:Threats-to-Validity}
Here, we review the threats to the validity of our proposed approach. We categorize the threats into three categories, namely internal, external, and dependability~\cite{SEKE_2010_Feldt_Validity-Threats}.
\subsection{Internal Validity}
Our results are based on the data set introduced by Mao et al.~\cite{ICST_2019_Mao_An-Extensive-Study-on-Cross-Project-Predictive-Mutation-Testing}. Therefore the values of independent variables computed by other researchers could affect our results. 

We assessed our approach based on metrics such as AUC, MCC, and Balanced Accuracy. However, there are a plethora of metrics for evaluating a machine learning model. We mitigated this threat by choosing metrics that have unbiased towards an imbalanced data set~\cite{EASE_2020_Yao_MCC-matters,ICPR_2010_Brodersen_Balanced-Accuracy}.

We exploited the Scott-Knott ESD test to compare our proposed approach to PMT(12) and PMT(95). The outcome of the test could have an impact on our results, yet, the Scott-Knott ESD test has been used in lots of software literature~\cite{ESE_2019_Kondo_The-impact-of-feature-reduction-techniques-on-defect-prediction-models,SANER_2019_Yu_An-empirical-study-of-learning-to-rank-techniques-for-effort-aware-defect-prediction,ESE_2017_Xia_What-do-developers-search-for-on-the-web,MSR_2017_Ghotra-A-large-scale-study-of-the-impact-of-feature-selection-techniques-on-defect,JSEP_2019_Catolino_An-extensive-evaluation-of-ensemble-techniques-for-software-change}.

\subsection{External Validity}
As for external validity, the data set we used is specifically gathered for Java projects. Therefore, it is not applicable to other programming languages.

In the evaluation part, we compared our results to the best PMT model, i.e., Random Forest. However, we also implemented other machine learning models in the previous papers~\cite{TSE_2019_Zhang_Predictive-Mutation-Testing,ICST_2019_Mao_An-Extensive-Study-on-Cross-Project-Predictive-Mutation-Testing}. To reduce this threat, we
implemented other machine learning models such as Logistic Regression and Deep Neural Networks. To save space, we avoid reporting their results in this paper, yet, they are available at our online appendix~\cite{PUMT}.
Note that, we also compared our proposed approach to the best machine learning model implemented in the previous research.

We also neglected equivalent mutants in the predicting process. Such mutants certainly would be survived since they are equivalent to the original program. However, finding equivalent mutants proved to be an undecidable problem~\cite{CA_1996_Offutt_Equivalent-Mutants,FSE_1998_Frankl_Empirical-Studies-Test-Effectiveness}.

\subsection{Dependability}
Reproducibility is a key factor in assessing a paper~\cite{IST_2018_Shepperd_The-role-and-value-of-replication}. Similar to Mao et al. and Zhang et al.~\cite{TSE_2019_Zhang_Predictive-Mutation-Testing,ICST_2019_Mao_An-Extensive-Study-on-Cross-Project-Predictive-Mutation-Testing}, we make our all source code publicly available for other researchers and practitioners to reproduce the results and further investigate the Predictive Mutation Testing~\cite{PUMT}. Predictive Mutation Testing is a new area, and there are lots of opportunities for other researchers to expand this dimension.
\section{Conclusions}\label{sec:Conclusions}
In this paper, we investigated the impact of uncovered mutants on the results of previous research regarding the Predictive Mutation Testing. Specifically, we found that such mutants can reduce drastically the performance of state-of-the-art techniques in terms of AUC. Then, we proposed an approach based on the combination of Random Forest and Gradient Boosting to handle that issue. Our results show that the proposed approach outperforms other techniques in terms of AUC, MCC, and Balanced Accuracy. Finally, we examined the most important features and found some contradiction when the impact of uncovered mutants was considered. For instance, \texttt{mmhalsteadDifficulty}, a  method-level Halstead difficulty metric, is the second most important feature in predicting the execution results of mutants in our proposed approach. However, this metric does not even appear on the most important features in other techniques. 

For future work, we are going to extend the data set to further improve the results and see the difference in the interpretation of our proposed approach. We also have a plan to add other metrics related to Abstract Syntax Tree (AST) to enrich the data set and study an impact of such metrics on PMT. Prior research demonstrated that observability metrics correlate with the mutation score~\cite{PeerJ_2019_Zhu_How-to-kill-them-all}. We are going to implement those metrics and examine the relationship between them and PMT.

\appendix
\section{Appendix: List of Features}
Here, we define the list of 30 features (borrowed from the Jhawk tool) used in the paper.
The \texttt{pp}, \texttt{cc}, and \texttt{mm} prefixes refer to package level, class level, and method level metrics. \tableref{Appendix} shows the 30 features as well as the definition of each one. 

\begin{table}[htpb]
  \caption{The list of features accompanied by definition~\cite{Jhawk}}
  \label{tab:Appendix}
  \begin{tabular}{ll}
   \toprule
    \bfseries Name&\bfseries Definition\\
   \midrule
   numExecuted & The number of times each mutant is executed\\
   MutatorClass & The type of the mutator operator\\ 
   numAssertInTC & The number of assertions in the test classes that execute the mutant\\
   numTestCover & The number of distinct test cases that cover the mutant\\
   ppavcc & Package level cyclomatic complexity\\
   cchalsteadCumulativeBugs& Class level cumulative Halstead bugs\\
   ppRVF & The total number of variables referred in the package\\
   ppnumberOfMethods & The number of methods each package has\\
   ppnumberOfClasses & The number of classes each package has\\
   ppmaintainabilityIndexNC & The package level maintainability index without considering comments\\
   ppfanout & The package level fan out\\
   ccmaintainabilityIndex & The class level maintainability index\\
   mmhalsteadDifficulty & The method level Halstead difficulty\\
   ppabstractness & The package level abstractness\\
   ppmaintainabilityIndex & The package level maintainability index\\
   ccexternalMethodCalls & The total number of method invocation in the class\\
   mminstanceVariablesReferenced~~&The number of fields referred in the method\\
   ccimportedPackages & The number of packages imported in the class\\
   ppdistance & The package level distance metric\\
   returnType & The type of the return value of the method\\
   ccfanIn & The class level fan in\\
   ppfanin & The package level fan in\\
   pploc & The package level lines of code\\
   ccmaintainabilityIndexNC & The class level maintainability index without considering comments\\
   mmexternalMethodsCalled & The number of method invocation in the method\\
   ppinstability & The package level instability\\
   ppmaxcc & Maximum cyclomatic complexity value in the package \\
   mmvariablesReferenced & The number of variables referred in the method\\
   ccunweightedClassSize & The sum of the number of methods and fields\\
   \bottomrule
  \end{tabular}
\end{table}

\clearpage

% Loading bibliography database
\bibliography{refs}

\end{document}